\documentclass[journal]{IEEEtran}

\usepackage{soul, color}
\sethlcolor{yellow}

\usepackage{cite}
\usepackage{textcomp}

\usepackage{graphicx}
\ifCLASSINFOpdf
\else
\fi
\usepackage{amsmath}
\usepackage{multirow}

\begin{document}
%
% paper title
% Titles are generally capitalized except for words such as a, an, and, as,
% at, but, by, for, in, nor, of, on, or, the, to and up, which are usually
% not capitalized unless they are the first or last word of the title.
% Linebreaks \\ can be used within to get better formatting as desired.
% Do not put math or special symbols in the title.
\title{A Lightweight Solution of Industrial Computed Tomography with Convolutional Neural Network}
%
%
% author names and IEEE memberships
% note positions of commas and nonbreaking spaces ( ~ ) LaTeX will not break
% a structure at a ~ so this keeps an author's name from being broken across
% two lines.
% use \thanks{} to gain access to the first footnote area
% a separate \thanks must be used for each paragraph as LaTeX2e's \thanks
% was not built to handle multiple paragraphs
%

\author{Guogang~Zhu, Jian~Fu ~\IEEEmembership{Member ~IEEE}% <-this % stops a space
\thanks{This work was supported in part by National Natural Science Foundation
	of China (51975026), the Joint Fund of Research Utilizing
	Large-scale Scientific Facilities by National Natural Science Foundation of
	China and Chinese Academy of Science (U1932111) and National Science and Technology Major
	Project of China (2018ZX04018001-006).
	}% <-this % stops a space
\thanks{Guogang Zhu is with School of Mechanical Engineering and Automation, Beihang University, 100191 Beijing, China.
    Jian Fu is with School of Mechanical Engineering and Automation, Beihang, 100191 Beijing, China,
	and Jiangxi Research Institute, Beihang University, 33000, Nanchang, China.  Email: fujian706@buaa.edu.cn, buaa\_zgg@buaa.edu.cn}}% <-this % stops a space

\maketitle

% As a general rule, do not put math, special symbols or citations
% in the abstract or keywords.
\begin{abstract}
As an advanced non-destructive testing and quality control technique, industrial computed tomography (ICT) has found many applications in smart manufacturing. The existing ICT devices are usually bulky and involve mass data processing and transmission. It results in a low efficiency and cannot keep pace with smart manufacturing. In this paper, with the support from Internet of things (IoT) and convolutional neural network (CNN), we proposed a lightweight solution of ICT devices for smart manufacturing. It consists of efforts from two aspects: distributed hardware allocation and data reduction. At the first aspect, ICT devices are separated into four functional units: data acquisition, cloud storage, computing center and control terminals. They are distributed and interconnected by IoT. Only the data acquisition unit still remains in the production lines. This distribution not only slims the ICT device, but also permits the share of the same functional units. At the second aspect, in the data acquisition unit, sparse sampling strategy is adopted to reduce the raw data and singular value decomposition (SVD) is used to compress these data. They are then transmitted to the cloud storage. At the computing center, an ICT image reconstruction algorithm and a CNN are applied to these compressed sparse sampling data to obtain high quality CT images. The experiments with practical ICT data have been executed to demonstrate the validity of the proposed solution. The results indicate that this solution can achieve a drastic data reduction, a storage space save and an efficiency improvement without significant image degradation. The presented work has been helpful to push the applications of ICT in smart manufacturing.
\end{abstract}

% Note that keywords are not normally used for peerreview papers.

% For peer review papers, you can put extra information on the cover
% page as needed:
% \ifCLASSOPTIONpeerreview
% \begin{center} \bfseries EDICS Category: 3-BBND \end{center}
% \fi
%
% For peerreview papers, this IEEEtran command inserts a page break and
% creates the second title. It will be ignored for other modes.
\IEEEpeerreviewmaketitle

\section{Introduction}
% The very first letter is a 2 line initial drop letter followed
% by the rest of the first word in caps.
%
% form to use if the first word consists of a single letter:
% \IEEEPARstart{A}{demo} file is ....
%
% form to use if you need the single drop letter followed by
% normal text (unknown if ever used by the IEEE):
% \IEEEPARstart{A}{}demo file is ....
%
% Some journals put the first two words in caps:
% \IEEEPARstart{T}{his demo} file is ....
%
% Here we have the typical use of a "T" for an initial drop letter
% and "HIS" in caps to complete the first word.
\IEEEPARstart{I}{dustrial} computed tomography (ICT) has been widely applied to industries such as aerospace \cite{fu2016multi, fu2018multi}, auto \cite{barciewicz2018computed, zheng2017non} and microelectronic\cite{asadizanjani2017pcb}. It generates nondestructively the high spatial resolution three dimensional (3D) mapping images of objects and enables an evaluation on the internal structures. ICT has played a key role to control the product quality in many fields since 1970s.

Smart manufacturing has been recently interested in ICT and many applications have been reported \cite{bauza2018realization, zikmund2019computed, sagbas2019industrial, akdogan2019re}. Bauza et al. presented the realization of industry 4.0 with high speed ICT in high volume production \cite{bauza2018realization}. Sagbas et al adopted ICT to execute nondestructive inspection of additive manufactured parts \cite{sagbas2019industrial}. Akdogan  et al applied ICT to re-engineering of manufacturing parts \cite{akdogan2019re}. The ICT device is a complicated integrity and generally consists of system control, data acquisition, image reconstruction, data storage and drawback recognition. During the ICT scanning, mass X-ray projection and 3D image data is generated, processed and transmitted. Cone beam computed tomography (CBCT) with FDK (Feldkamp-Davis-Kress) \cite{feldkamp1984practical} has been a common used ICT mode since it keeps a good balance between efficiency and quality. Typically, the generated data of a CBCT scanning can reach 10GB. So ICT is bulky, quite time-consuming and inefficient. It currently cannot satisfy the demand from smart manufacturing.

Internet of Things (IoT) is one of the drivers and foundations of data-driven innovations in smart manufacturing \cite{yang2019internet}. It forms a global information network composed of a large number of interconnected "Things" (e.g., materials, sensors, equipments, people, products, and supply chain) \cite{yang2019internet}. By interconnecting these manufacturing things, IoT could achieve an effective digital integration of the entire manufacturing enterprise. Adopting IoT to separate and distribute the functional units can slim ICT device and let it be lightweight. Moreover, it can also make each functional unit more sophisticated and enable the share of the same functional units. However, the mass data transmission from ICT is a heavy burden and IoT cannot afford it. So it is necessary to reduce the ICT data to improve the transmission efficiency.

Sparse sampling is a good idea to reduce ICT data. However, ICT image reconstruction algorithms have a demand for the completeness of projection data. The ICT images suffer from artifacts since sparse sampling cannot meet this completeness condition. Developing the sparsity of the scanned objects, compress sensing can relax the completeness constraint and has been proved to be a powerful technique for sparse sampling image reconstruction \cite{hsieh2018compressed}. Total variation(TV)\cite{rong2014prior} has also shown its ability in preserving edges and suppressing artifacts for sparse sampling image reconstruction. These techniques perform better than the conventional analytical algorithms, but they encounter some limits such as expensive time consumption and the complicated parameter selection.

The sparse sampling data still contains redundant information. It can be further compressed to save the storage space and improve the data transmission efficiency. The redundance lies on the the similarities in the neighboring pixels\cite{kahu2013image}. Compared with the visible light images, ICT images have much more redundancy due to the similarities of materials and structures of the inspected objects. The solution to address this problem is to compress the projection while maintaining acceptable image quality. Singular value decomposition (SVD) in image compression has attracted more and more interests \cite{prasantha2007image}. SVD aims to find the low-rank approximation of the image matrix and represent it with much less data.

Sparse sampling and SVD can greatly reduce the ICT data, but they may lead to a degradation in ICT reconstruction images. Deep learning(DL) has been recently more and more popular and provides a possible solution for this degradation. As a typical DL technique, convolutional neural network(CNN) has been widely applied to image processing such as segmentation \cite{Unet}, super resolution\cite{yu2018wide} and  restoration\cite{zhang2020residual}. It has also found many applications in CT. Han et al reported a deep learning residual architecture for sparse sampling CT \cite{han2016deep}. They first estimated the artifacts caused by the sparse sampling and then subtracted it from the CT image to obtain the artifact-free image. Jin et al developed a deep CNN for inverse problems in CT \cite{jin2017deep}. They combined multi-resolution decomposition with residual learning to remove the artifacts. Pelt et al proposed a mixed-scale dense CNN for image analysis \cite{pelt2018mixed}.They adopted dilated convolutions to capture features at different image scales and densely connected all feature maps with each other. This architecture reduces parameters and avoids overfitting. Zhang et al \cite{zhang2018sparse} presented a sparse sampling CT reconstruction method by combining DenseNet \cite{huang2017densely} and deconvolution to remove the artifacts. Dong et al \cite{dong2019deep} reported a deep learning framework for CT with sparse-view projections. It is based on U-Net to estimate the complete projections from sparse-view projections. The above mentioned methods are based on 2D fan beam CT. For 3D CBCT, Jiang et al \cite{jiang2019augmentation} proposed a symmetric residual convolutional neural network (SR-CNN). This architecture is able to reserve edges and structures in under-sampled CBCT images. Yang et al \cite{yang2020streaking} adopted residual leaning network to suppress the streaking artifacts in sparse sampling CBCT and made a great progress.

In this paper, supported by IoT and CNN, we proposed a lightweight solution of ICT for smart manufacturing. It consists of efforts from two aspects: distributed hardware allocation and data reduction. First, the ICT device in the production line is simplified to be a data acquisition unit and all the other units are separated and distributed by IoT. It not only slims the ICT device, but also permits ICT devices from different production lines can share the same functional units. Moreover, the functional units will obtain space to develop to be more sophisticated and powerful. Second, sparse sampling strategy is adopted in the data acquisition unit to reduce the raw data and SVD is adopted to compress the raw data. The compressed data is then upload to the shared and distributed ICT units for further processing and storage. When image reconstruction is executed, the raw projection data is first recovered from the compressed data and reconstructed with FDK algorithm to generate the original CT images. A dense connection based CNN is then applied to the original CT images to remove the artifacts caused by sparse sampling and SVD. This CNN originates from the framework proposed by our group \cite{fu2019deep} and can reduce the number of parameters and achieve a high learning efficiency. In this article, we focus on data reduction. The distributed hardware allocation depends on IoT and will not be discussed in detail. The experiments with practical ICT data have been executed to demonstrate the validity of the proposed solution. The results indicate that this technique can achieve a drastic data reduction and an efficiency improvement without significant image degradation.

Briefly, the contributions in this article are as follow:\par
1) Present a lightweight solution of ICT for smart manufacturing.\par
2) Adopt distributed hardware allocation to slim ICT devices and permit the share of the same functional units.\par
3) Adopt sparse sampling strategy and SVD to reduce data. \par
4) Construct a CNN to deal with ICT image degradation. \par

%\hfill mds

%\hfill August 26, 2015

\section{Methods}
% needed in second column of first page if using \IEEEpubid
%\IEEEpubidadjcol

\subsection{Solution Overview}
\begin{figure*}[!t]
\centering
\includegraphics[width=7in]{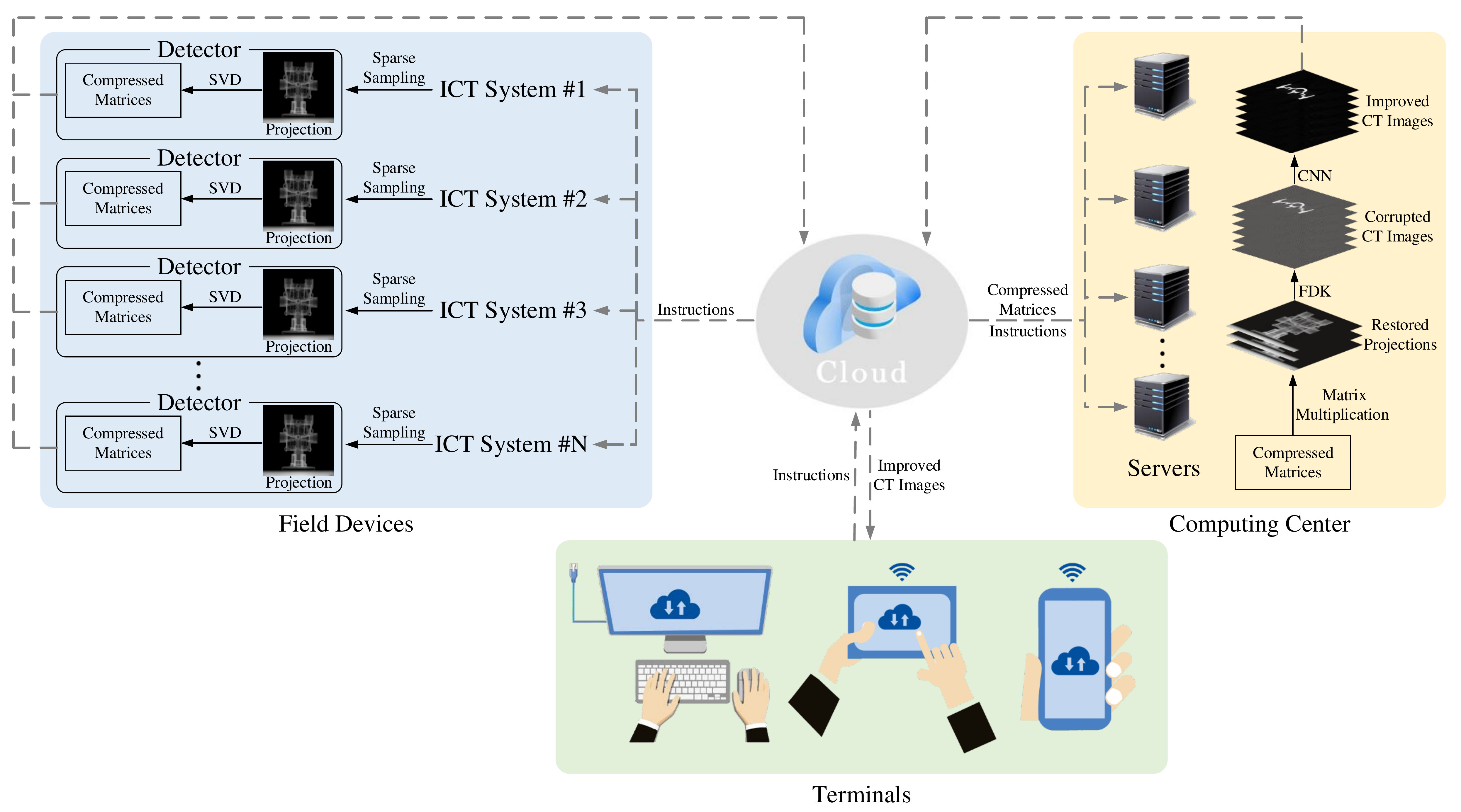}
\caption{Lightweight solution of ICT for smart manufacturing. It consists of four parts: field devices, cloud storage, computing center and terminals. They together implement distributed allocation of ICT functional units and data reduction. }
\label{solution}
\end{figure*}

As illustrated in Fig.\ref{solution}, the proposed solution consists of four parts: field ICT devices, cloud storage, computing center and terminals. The cloud storage acts as a hub through which data and instructions are transmitted among other three parts. The field ICT devices are simplified to be data acquisition units and from different production lines. They receive the instructions from terminals and make response. They acquire, compress and transmit the compressed data to cloud storage. Other ICT units such as image reconstruction and drawbacks recognition are centralized to computing center. Terminals may be industrial computers or portable mobile devices such as smart phones and tablets. They control field devices and computing center, and have access to the ICT projection and image data saved at the cloud storage.

The field devices adopt sparse sampling strategy and SVD technique to reduce the raw projection data. Much less data is acquired in sparse sampling and the data acquisition time also greatly decreases. SVD is a linear matrix transformation used for image compression to remove the data redundancy. By applying SVD to the sparse sampling data and discarding lower singular values, the data could be approximated with three small size low-rank matrices and a great compression is obtained. It also saves the storage space. SVD can be implemented by hardware and easily embedded in the detector.

The computing center consists of a number of servers. They receive instructions from terminals and data from cloud storage and upload the improved CT images to cloud storage. The computing center performs three tasks: compressed projection restoration, image reconstruction and image improvement. First, the SVD compressed projections are restored. Then, FDK algorithm is applied to these restored projections to obtain the corrupted CT images. Finally, a CNN is applied to the corrupted CT images to remove the artifacts caused by sparse sampling and SVD compression. This CNN is based on densely connected pattern and U-Net. It keeps a good balance between the number of parameters and the performance.

This solution in Fig.\ref{solution} implements distributed allocation of ICT functional units and data reduction. By distributed hardware allocation, ICT devices become slim and lightweight and the same functional units can be shared. Data reduction not only improves the transmission efficiency, but also save the storage space. So it provides a possible lightweight ICT solution for smart manufacturing.

\subsection{Neural Network}
\begin{figure}[!t]
	\centering
	\includegraphics[width=3.5in]{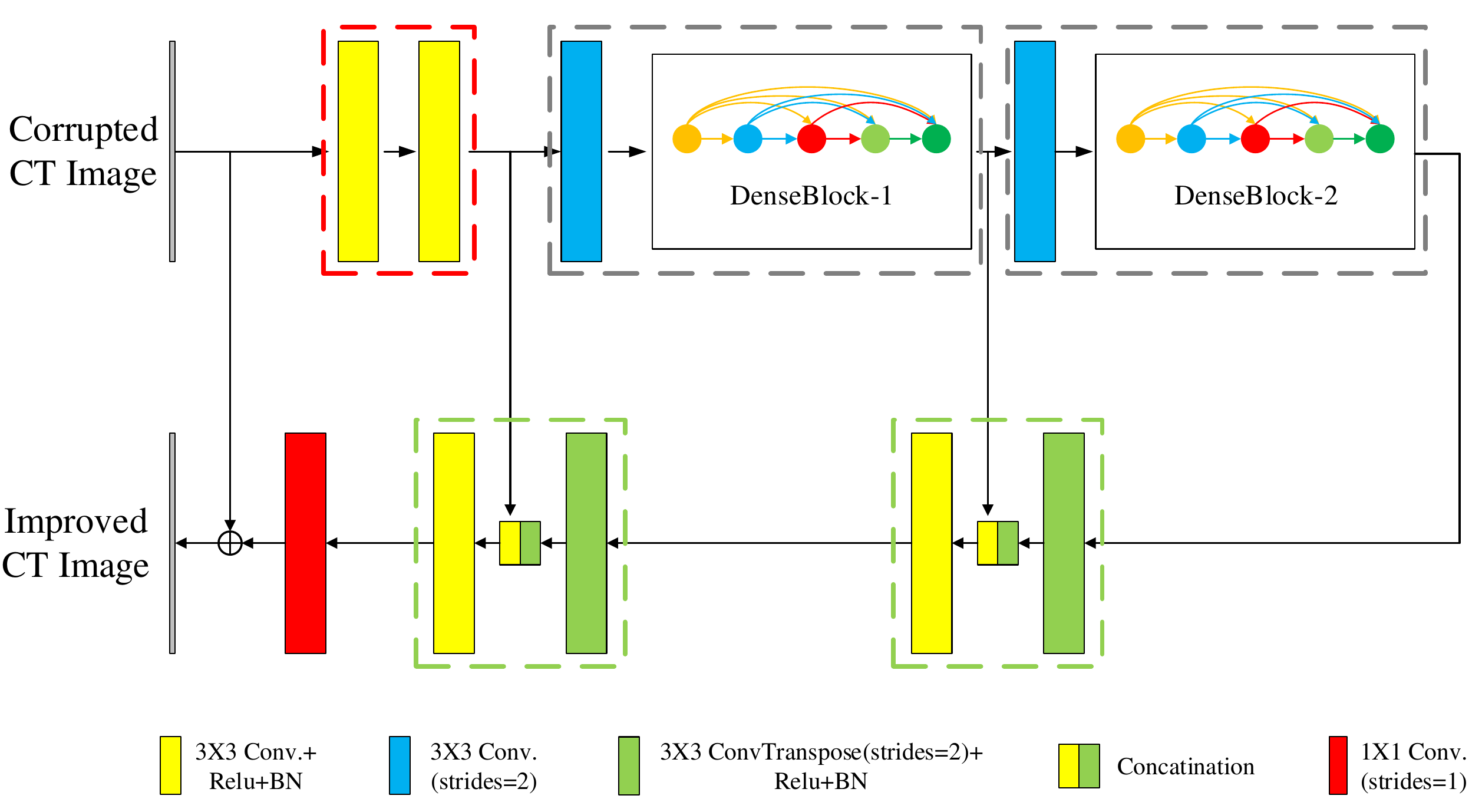}
	\caption{The architecture of the CNN used in this paper. It takes fully advantages of U-Net and DenseNet.}
	\label{cnn_architecture}
\end{figure}

Depicted in Fig.\ref{cnn_architecture}, the architecture of the CNN in Fig.\ref{solution} is modified from the framework proposed by our group \cite{fu2019deep}. It takes advantages of U-Net \cite{Unet} and DenseNet\cite{huang2017densely} to remove the artifacts caused by sparse sampling and SVD compression. As shown in Fig.\ref{cnn_architecture}, the corrupted CT images are fed into the network and improved CT images are obtained. It consists of four kinds of function units as follow:\par
i)Two convolutional layers indicated by the two yellow solid rectangles surrounded by a red dashed line which are used to extract primary features.\par
ii)Two components surrounded by two grey dashed lines which contain down-sampling layers and DenseBlocks. The down-sampling is realized with convolutional layer with a stride of 2. These two components act as encoders and obtain multi-scale features.\par
iii)Two components surrounded by two green dashed lines which are used for decoding. They restore high resolution features from the low resolution features and concatenate them with lower level features.\par
iv)A convolutional layer with a filter size of $1\times1$ indicated by the red solid rectangle. It compresses the stacked feature layers into one layer to keep the same size with the fed corrupted image. \par

As an example, the parameters of the network are listed in the Table \ref{network_para} where the size of the fed corrupted CT image is $700 \times 700$ pixels.

\begin{table}[!t]
	% increase table row spacing, adjust to taste
	%\renewcommand{\arraystretch}{1.3}
	% if using array.sty, it might be a good idea to tweak the value of
	% \extrarowheight as needed to properly center the text within the cells
	\caption{Parameters of the Conventional Neural Network in Fig.\ref{cnn_architecture} }
	\label{network_para}
	\centering
	% Some packages, such as MDW tools, offer better commands for making tables
	% than the plain LaTeX2e tabular which is used here.
	\begin{tabular}{c|c|c}
		\hline
		\textbf{Layers} & \textbf{Input Size} & \textbf{Ouput Size} \\ \hline\hline
		\begin{tabular}[c]{@{}c@{}}$3\times3$ Conv.(strides=1)\\-Relu-BN\end{tabular} & $700\times700\times1$ & $700\times700\times32$ \\ \hline
		
		\begin{tabular}[c]{@{}c@{}}$3\times3$ Conv.(strides=1)\\-Relu-BN\end{tabular} & $700\times700\times32$ & $700\times700\times32$ \\ \hline
		
		$3\times3$ Conv.(strides=2) & $700\times700\times32$ & $350\times350\times32$ \\ \hline
		
		DenseBlock-1 & $350\times350\times32$ & $350\times350\times160$ \\ \hline
		
		$3\times3$ Conv.(strides=2) & $350\times350\times160$ & $175\times175\times32$ \\ \hline
		
		DenseBlock-2 & $175\times175\times32$ & $175\times175\times160$ \\ \hline
		
		\begin{tabular}[c]{@{}c@{}}$3\times3$ ConvTrans.(strides=2)\\-Relu-BN\end{tabular} & $175\times175\times160$ & $350\times350\times160$ \\ \hline
		
		\multirow{2}{*}{Concatenation} & $350\times350\times160$ & \multirow{2}{*}{$350\times350\times320$} \\ \cline{2-2}
		& $350\times350\times160$ &  \\ \hline
		
		\begin{tabular}[c]{@{}c@{}}$3\times3$ Conv.(strides=1)\\-Relu-BN\end{tabular} & $350\times350\times320$ & $350\times350\times160$ \\ \hline
		
		\begin{tabular}[c]{@{}c@{}}$3\times3$ ConvTrans.(strides=2)\\-Relu-BN\end{tabular} & $350\times350\times160$ & $700\times700\times160$ \\ \hline
		
		\multirow{2}{*}{Concatenation} & $700\times700\times160$ & \multirow{2}{*}{$700\times700\times192$} \\ \cline{2-2}
		& $700\times700\times32$ &  \\ \hline
		
		\begin{tabular}[c]{@{}c@{}}$3\times3$ Conv.(strides=1)\\-Relu-BN\end{tabular} & $700\times700\times192$ & $700\times700\times96$ \\ \hline
		
		$1\times1$ Conv.(strides=1) & $700\times700\times96$ & $700\times700\times1$ \\ \hline
		
		\multirow{2}{*}{Add} & $700\times700\times1$ & \multirow{2}{*}{$700\times700\times1$} \\ \cline{2-2}
		& $700\times700\times1$ &  \\ \hline
		
	\end{tabular}
\end{table}

\subsubsection{Skip Connection}
The CNN in Fig.\ref{cnn_architecture} contains two kinds of skip connections: element-wise addition between input and output in the form of residual learning \cite{he2016deep} and feature concatenation among different stages.\par

The element-wise addition is formulated in (\ref{res_learning}), where $X$ represents the fed image, $H()$ the nonlinear mapping of the network and $\hat{Y}$ the output of the network. This operation forces the network to recognize and remove the artifacts from the corrupted CT images, rather than remove the artifacts and build up the whole CT images at the same time. It simplifies the learning process.

\begin{equation}
	\hat{Y} = H(X) + X
	\label{res_learning}
\end{equation}

Another skip connection is feature concatenation among different stages. It aims to integrate different levels of features. It is well known that DL naturally takes into account low, middle and higher level features. In the CNN in Fig.\ref{cnn_architecture}, the high level features that contain wealth semantic information are hard to recover the spatial information. On the contrary, the low level features that contain less semantic information retain a good spatial accuracy. So the low level features are concatenated with high-level features to ensure a best result.

\subsubsection{DenseBlock}
\begin{figure}[!t]
	\centering
	\includegraphics[width=3.5in]{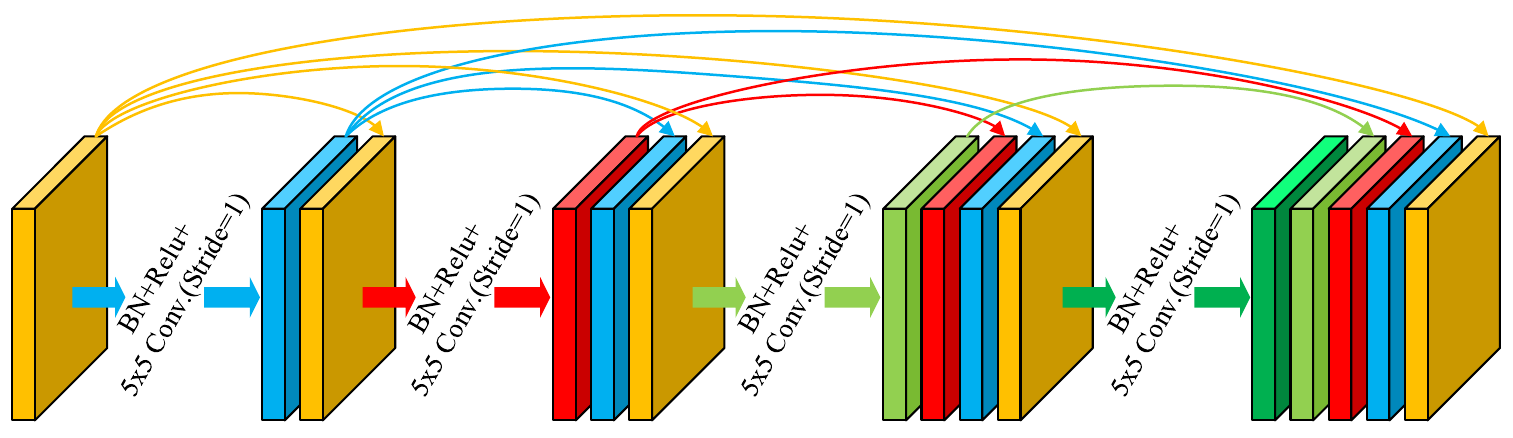}
	\caption{The structure of DenseBlock, each layer gets inputs from all preceding layers and passes its own features to all subsequent layers.}
	\label{DenseBlock}
\end{figure}

In the proposed solution, the CNN should be lightweight to obtain a high processing speed. We adopt DenseBlock technique to reduce the parameters. The structure of DenseBlock is shown in Fig.\ref{DenseBlock}. Each layer takes all preceding layers as input and passes its own features to all subsequent layers. The DenseBlock in Fig.\ref{DenseBlock} contains four BN-Relu-$5\times5$ Conv.(strides=1) layers. Compared with the preceding layer, each layer produces $k$ new features. Features from all the preceding layers are reused through dense connections. It could explore the full potential of the network and yield condensed models with higher parameter efficiency. Some researches have shown that a small number of features are sufficient for DenseBlock to obtain acceptable results. Densely connection discards redundant features and focuses on taking advantage of retained features, which dramatically reduces the number of parameters and improves the processing speed.\par

Moreover, densely connection makes the network easier to train since each layer in the same block can directly access the gradients from the cost function and the original input. It results in an implicit deep supervision \cite{huang2017densely}. Additionally, each layer generates only $k$ new features. The contribution to the final result from each one layer is controlled. It will overcome the problem of overfitting in a small batch training set.

The parameters of the DenseBlock in Fig.\ref{DenseBlock} are listed in the Table \ref{denseblock_para}, where $H$ and $W$ represent the height and width of the features respectively.

\begin{table}[!t]
	% increase table row spacing, adjust to taste
	%\renewcommand{\arraystretch}{1.3}
	% if using array.sty, it might be a good idea to tweak the value of
	% \extrarowheight as needed to properly center the text within the cells
	\caption{Parameters of Denseblock in Fig.\ref{DenseBlock}}
	\label{denseblock_para}
	\centering
	% Some packages, such as MDW tools, offer better commands for making tables
	% than the plain LaTeX2e tabular which is used here.
	\begin{tabular}{cc|c|c}
		\hline
		& \textbf{Layers} & \textbf{Input Size} & \textbf{Output Size} \\ \hline\hline
				
		\multicolumn{1}{c|}{\multirow{3}{*}{1}} & BN-Relu-$5\times5$ Conv. & $H\times W\times V$ & $H\times W\times k$ \\ \cline{2-4}
		\multicolumn{1}{c|}{} & \multirow{2}{*}{Concatenation} & $H\times W\times V$ & \multirow{2}{*}{$H\times W\times (V+k)$} \\ \cline{3-3}
		\multicolumn{1}{c|}{} &  & $H\times W\times k$ &  \\ \hline

		\multicolumn{1}{c|}{\multirow{3}{*}{2}} & BN-Relu-$5\times5$ Conv. & $H\times W\times (V+k)$ & $H\times W\times k$ \\ \cline{2-4}
		\multicolumn{1}{c|}{} & \multirow{2}{*}{Concatenation} & $H\times W\times (V+k)$ & \multirow{2}{*}{$H\times W\times (V+2k)$} \\ \cline{3-3}
		\multicolumn{1}{c|}{} &  & $H\times W\times k$ &  \\ \hline
		
		\multicolumn{1}{c|}{\multirow{3}{*}{3}} & BN-Relu-$5\times5$ Conv. & $H\times W\times (V+2k)$ & $H\times W\times k$ \\ \cline{2-4}
		\multicolumn{1}{c|}{} & \multirow{2}{*}{Concatenation} & $H\times W\times (V+2k)$ & \multirow{2}{*}{$H\times W\times (V+3k)$} \\ \cline{3-3}
		\multicolumn{1}{c|}{} &  & $H\times W\times k$ &  \\ \hline
		
		\multicolumn{1}{c|}{\multirow{3}{*}{4}} & BN-Relu-$5\times5$ Conv. & $H\times W\times (V+3k)$ & $H\times W\times k$ \\ \cline{2-4}
		\multicolumn{1}{c|}{} & \multirow{2}{*}{Concatenation} & $H\times W\times (V+3k)$ & \multirow{2}{*}{$H\times W\times (V+4k)$} \\ \cline{3-3}
		\multicolumn{1}{c|}{} &  & $H\times W\times k$ &  \\ \hline
		
	\end{tabular}
\end{table}

\subsubsection{Network Training}
The network training is executed according to the following steps:\par
i)A set of corrupted CT images are matched with the corresponding ground truth into many pairs of training data. Each pair of training data includes a corrupted CT image obtained from restored projection and its corresponding ground truth obtained from the complete projection.\par
ii)These training data are fed into the CNN depicted in Fig.\ref{cnn_architecture} one pair by one pair to optimize the network.\par
iii)Repeat steps i) and ii) until the learning converges.\par

After training, the CNN is determined and ready to be used to obtain high quality CT images from the corrupted ones.

\subsection{Singular Value Decomposition}
The SVD of the image matrix $\mathbf{G} \in \mathbf{R}_{m \times n}$ is formulated in (\ref{SVD}). In this equation, $\mathbf{U}=(\mathbf{u}_1,\dots, \mathbf{u}_m)\in\mathbf{R}^{m\times m}$ and $\mathbf{V}=(\mathbf{v}_1,\dots, \mathbf{v}_n)\in\mathbf{R}^{n\times n}$ are the matrices with orthonormal columns.  $\mathbf{\Sigma}=diag(\sigma_1, \sigma_2, \dots, \sigma_N) \in \mathbf{R}_{m \times n}$ is the diagonal matrix. The vectors $\mathbf{u}_i$ and $\mathbf{v}_i$ are the left and right singular vectors of $\mathbf{G}$. The diagonal entries $\sigma_i$ of $\mathbf{\Sigma}$ are singular values of $\mathbf{G}$. As shown in (\ref{sigular_value}), they are no less than zero and sorted by value descending.\par

\begin{equation}
\label{SVD}
	\mathbf{G} = \mathbf{U}\mathbf{\Sigma}\mathbf{V}^{T}
	=\sum_{i=1}^{N}\sigma_{i}\mathbf{u}_{i}\mathbf{v}_{i}^{T}
\end{equation}

\begin{equation}
\label{sigular_value}
	\sigma _1\ge\sigma _2\ge\cdots\ge\sigma _N\ge 0
\end{equation}

Lower singular values contain negligible image information which can be discarded without significant image degradation. Thus, the image matrix $\mathbf{G}$ could be approximated by the first $k$ singular values, as shown in (\ref{SVDrecon}). This approximation will reduce the storage space from $m \times n$ bytes to $k\times (m+n+1)$ bytes. The compression ratio (CR) given by (\ref{compress_ratio}) can be used to evaluate the compression performance.\par

\begin{equation}
\label{SVDrecon}
\mathbf{G} \approx \sum_{i=1}^{k}\sigma_{i}\mathbf{u}_{i}\mathbf{v}_{i}^{T}
\end{equation}

\begin{equation}
\label{compress_ratio}
CR_{SVD}=\frac{m\times n}{k\times (m+n+1)}
\end{equation}

\begin{figure}[!t]
	\centering
	\includegraphics[width=3.5in]{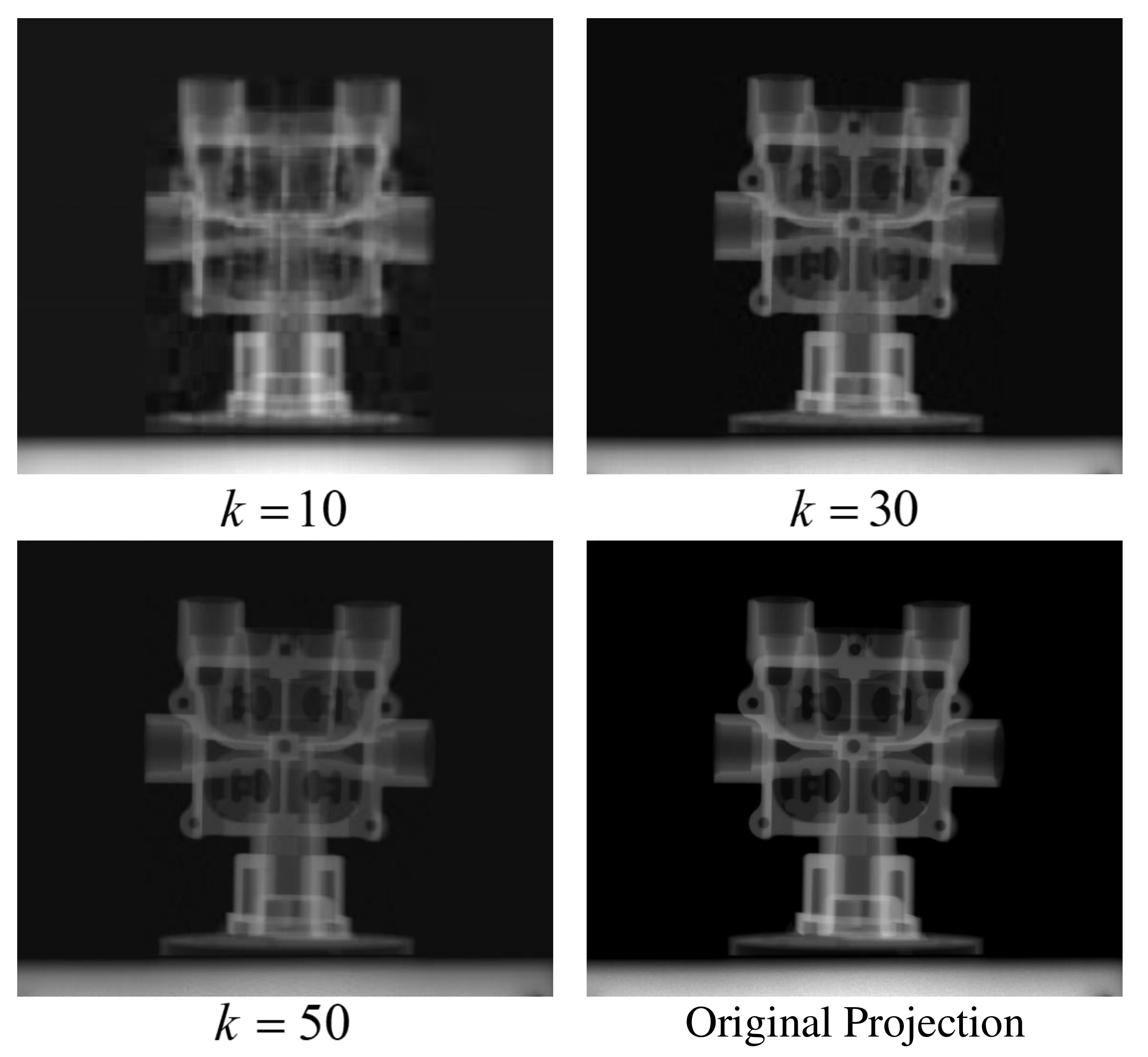}
	\caption{Compressed projections with different number of singular values.}
	\label{SVD_Compared}
\end{figure}

\begin{figure}[!t]
	\centering
	\includegraphics[width=3.5in]{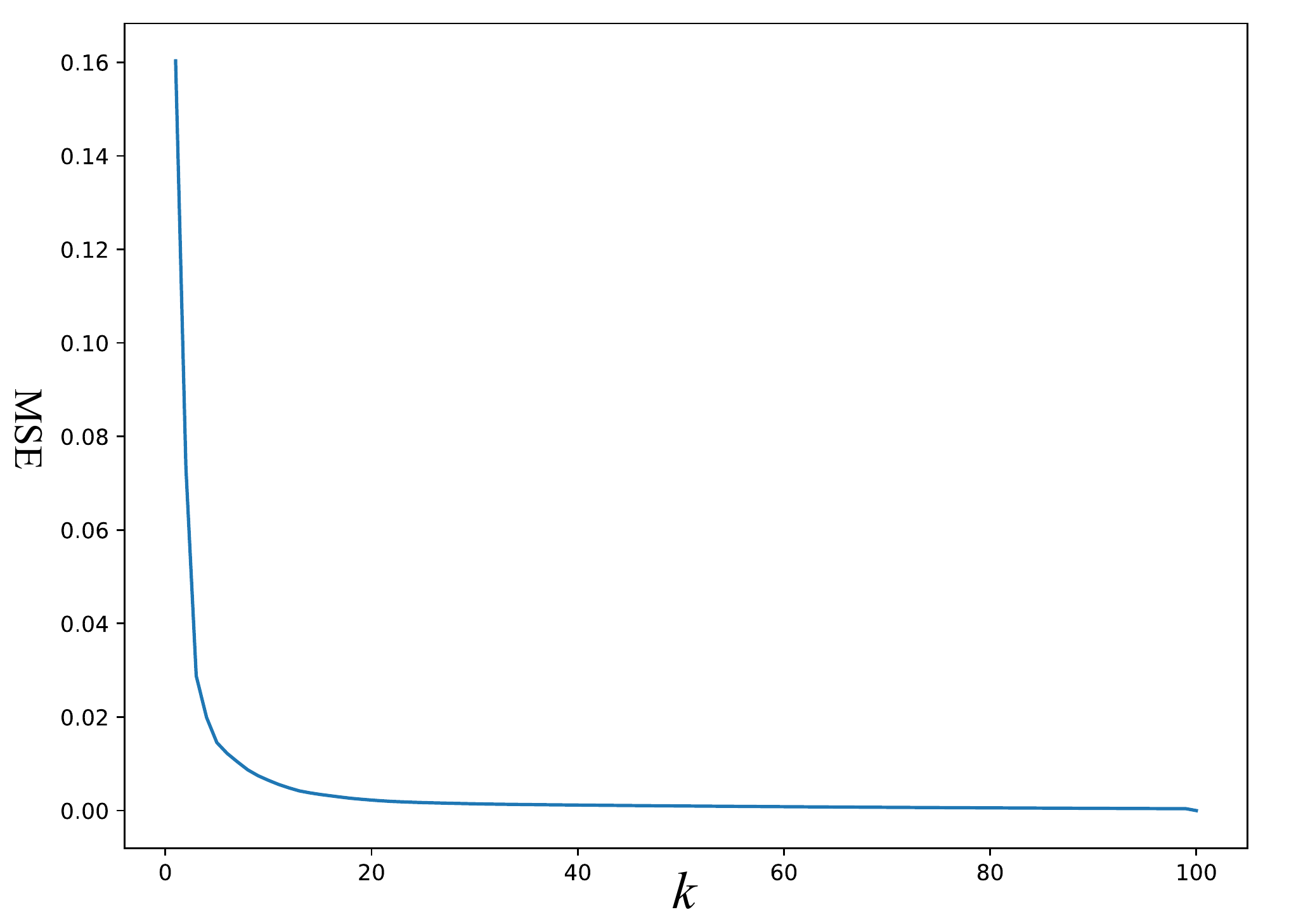}
	\caption{The MSE curve between the compressed projections and the original one. $k$ represents the number of singular values in SVD compression.}
	\label{mse_k}
\end{figure}

In ICT engineering applications, many objects are homogeneous and the internal structures are similar each other. Thus, the ICT projection data contains much more redundant information that is suitable for the compression with SVD. As an example, Fig.\ref{SVD_Compared} shows an ICT projection and its three compressed projections with different number of singular values. Fig.\ref{mse_k} shows the curve of mean square error(MSE) between the original ICT projection and its compressed one with SVD. Along with the increase of value of $k$, the MSE value descends rapidly. After the value of $k$ reaches 30, the MSE value has no visible change. It indicates that ICT projection can be approximated with the first singular values.

\subsection{FDK Reconstruction Algorithm}
FDK is a CBCT reconstruction algorithm \cite{feldkamp1984practical}. It has been the dominated technique in practical CBCT reconstruction. FDK algorithm has many extension versions and the one for a planar detector consists of the following three steps:\par

i)As shown in (\ref{fdk_weight}), weight the projection data. In this equation, $p_{\beta}(a,b)$ represents the original projection data acquired by the detector element at the $a_{th} $ row and $b_{th}$ column under the sampling angle $\beta$ .
\begin{equation}
\label{fdk_weight}
p_{\beta}^{'}(a,b)=p_{\beta}(a,b)\frac{R}{\sqrt{R^{2} + a^{2} + b^{2}}}
\end{equation}

ii)As shown in (\ref{fdk_filter}), execute one-dimensional filtering on the weighted data in (\ref{fdk_weight}). In this equation, $h(a)$ represents the ramp filter.
\begin{equation}
\label{fdk_filter}
\tilde{p}_{\beta}(a,b)=p_{\beta}^{'}(a,b)\ast h(a)
\end{equation}

iii)As shown in (\ref{fdk_bp}), do back projection reconstruction. In this equation, $f(x,y,z)$ represents the reconstruction volume and $R$ is the distance from X-ray source to detector.
\begin{equation}
\label{fdk_bp}
f(x,y,z)=\int_{0}^{2\pi} \frac{R^2}{U_{\beta}(x,y)}\tilde{p}_{\beta}(a,b)d\beta
\end{equation}

\begin{equation}
\label{U}
U_{\beta}(x,y)=R + x\cos\beta + y\sin\beta
\end{equation}

FDK algorithm has high reconstruction speed and accuracy when the projection data is complete. However, dramatic image degradation will occur when it is applied to sparse sampling projection data.

\section{Experiments}
\subsection{Data Preparation}
The experimental data set is prepared by executing the CBCT scanning of two aluminum auto components, No.1 and No.2, with the ICT system developed by our group. These two components have different shapes and internal structures. The data set of the NO.1 component is for training and the data set of the NO.2 component is for testing.

The used ICT system plays as a data acquisition unit and consists of an 160 kV X-ray source, YTU160-D01, from YXLON company (Germany) and a 2D X-ray flat panel detector, XRD 1621, from PerkinElmer company (USA). The X-Ray source works with a tube voltage 120$keV$ and a tube current 2.0$mA$. The detector has an element size of 0.2$mm$ and works in an internal trigger mode. The field of view has a size of about $410\times410 mm^2$.

The CBCT scanning angular increment is $0.5^\circ$ and totally 720 projections are acquired within $360^\circ$. At each scanning angular position, a 2D projection image with a size of $2048\times1716$ pixels is recorded by the 2D X-ray flat panel detector. So for each component, the projection data set has a size of $720\times2048\times1716$ pixels and occupies a storage space of 9.4263GB. It is complete for CT image reconstruction. Fig.\ref{SVD_Compared} shows one of the 2D projection images of the No.1 component.

The sparse sampling data set is obtained by extracting some projections from the complete data set. This extraction is with an equal angular interval named sparse sampling factor. In these experiments, the sparse sampling factor is set to be 12. The spare sampling CBCT projection data set has a size of $60\times2048\times1716$ pixels and occupies a storage space of 0.7855GB.

According to the SVD compression performance curve in Fig.\ref{mse_k}, the first 30 singular values are retained. With the sampling sparse factors 12, the compressed sparse sampling data set has a size of $60 \times (30 \times 2048 + 30 + 30 \times 1716)$ and just need a storage space of 0.0252GB. They will be transferred from the field ICT device to the cloud storage and fed to the computing center.

The 3D ICT volume image of each auto component consists of 460 2D cross section slice images with a size of $460\times700\times700$ pixels. They are generated by applying FDK algorithm to the projection data set at the computing center. The results from the complete projection data set with a size of $720\times2048\times1716$ pixels are treated as the ground truth.

Considering the spatial structure similarity of neighboring 2D cross section slices, in order to improve the learning efficiency and accuracy, only one of the neighboring 9 slices is selected for training. Totally, 50 slice images are selected from the 460 CT images of the No.1 auto component for training. We adopt some image transform operations such as rotation and translation to expand the training data set from 50 images to 550 images. All the 460 ICT slice images of the No.2 component are used for testing.\par

\subsection{Implementation}
The CNN is implemented with Python 3.7.3 and Tensorflow 1.14.0. It runs on a workstation with a CPU Intel(R) Core(TM) i7-7700 HQ and a GPU Nvidia GTX 1060 with Max-Q Design 6GBytes.

As shown in (\ref{mse}), the loss function is based on the mean square error (MSE), where $\hat{Y}$ denotes the output image and $Y$ represents the label image.
\par\begin{equation}
loss = \Vert Y - \hat{Y} \Vert_{2}^{2}
\label{mse}
\end{equation}

The network parameters are initialized using a Gaussian distribution with a mean value of zero and a standard deviation of $\sqrt{\frac{2}{n_{in}}}$, in which $n_{in}$ indicates the number of input units in each layer.\par

Adaptive momentum estimation (Adam) optimizer algorithm \cite{kingma2014adam} is adopted to train the network. The initial learning rate is $1\times10^{-4}$ and gradually reduced to $1\times10^{-6}$. All the models are trained for 100 epochs. It takes about 14 hours to complete the training. During testing, it takes about 0.8s to obtain an improved CT image. \par

\subsection{Quantitative Evaluation}
The performance of the proposed solution need to be evaluated from two aspects: image quality and compression ratio.\par

For image quality evaluation, we select feature similarity (FSIM) \cite{zhang2011fsim} and information weighted SSIM (IW-SSIM) \cite{wang2010information} as the indicators according to the research of Zhang et al\cite{zhang2012comprehensive}.\par

The total compression ratio of this solution is calculated with (\ref{Total_CR}). Here, $m$ and $n$ represent the height and width of projection, $k$ the number of retained singular values, $views$ the number of the sparse sampling projections.\par

\begin{equation}
	\begin{aligned}
		\label{Total_CR}
		CR &= CR_{SVD} \times \frac{720}{views}\\&=\frac{m\times n\times 720}{k\times (m+n+1)\times views}
	\end{aligned}
\end{equation}

\subsection{Results}
\begin{figure}[!t]
	\centering
	\includegraphics[width=3.5in]{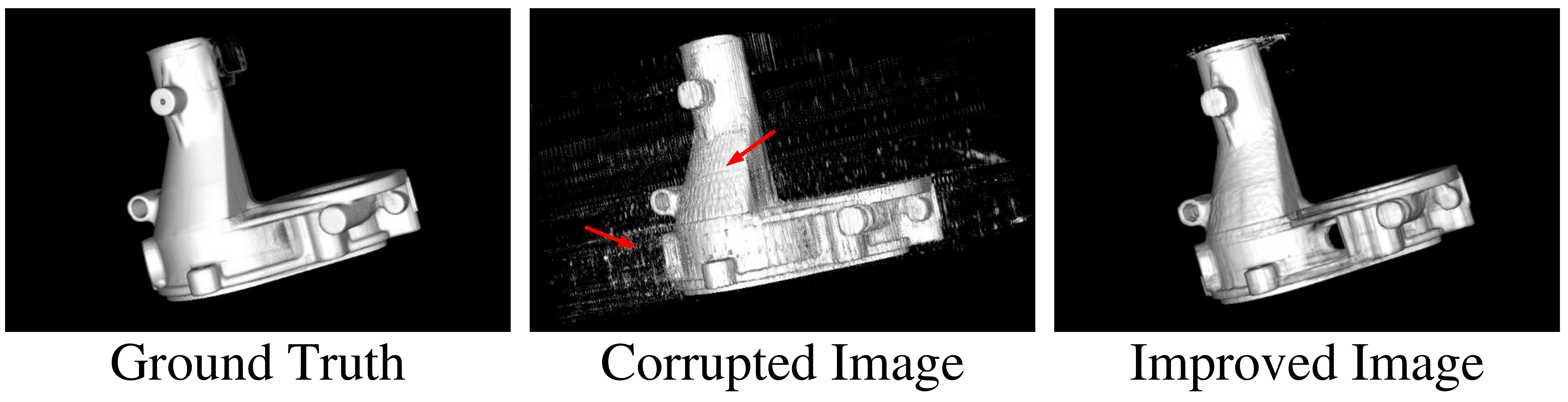}
	\caption{3D volume rendering of the 3D ICT images of the No.2 auto component with a sparse sampling factor 12 and SVD compression ratio 30. From left to right, the ground truth, the one before the improvement with CNN and the one after the improvement with CNN.}
	\label{result3d_60}
\end{figure}

Fig.\ref{result3d_60} presents the 3D ICT images of the No.2 auto component before and after the improvement with CNN. It corresponds to the case with a sparse sampling factor 12 and the first 30 singular values. Compared with the ground truth, the one before the improvement with CNN has coarse surfaces and artifacts, caused by the sparse sampling and SVD compression and indicated by the red arrows in Fig.\ref{result3d_60}. In the contrary, the one after the improvement with CNN has no visible artifacts.

\begin{figure}[!t]
	\centering
	\includegraphics[width=3.5in]{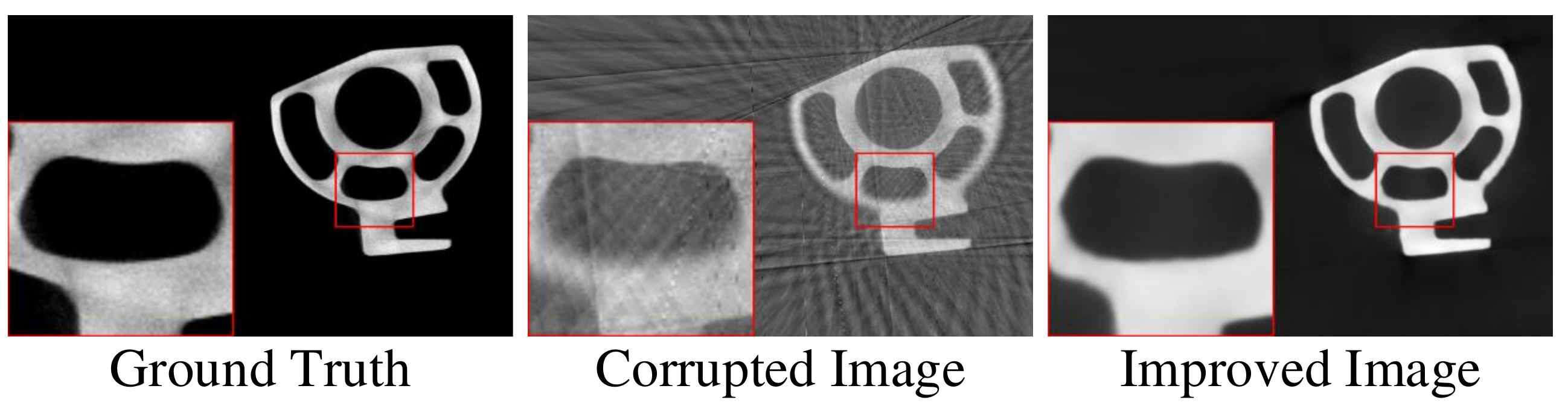}
	\caption{One typical 2D cross section slice of the 3D ICT image in Fig. \ref{result3d_60}. Indicated by the red box, a region of interest (ROI) is zoomed in for better visual observation and quantitative evaluation.}
	\label{result60}
\end{figure}

Shown in Fig.\ref{result60}, one typical 2D cross section slice is selected from the 3D ICT image and a region of interest (ROI) is zoomed in for better visual observation and quantitative evaluation. As observed in the 3D images, the streak artifacts are quite visible in the 2D slices before the improvement with CNN. After the processing, the streak artifacts disappear from the 2D slice images.

Table \ref{table_imagevalue} lists the FSIM and IW-SSIM values of presented images in Fig.\ref{result60}. Table \ref{table_evaluation} lists the averaged FSIM and IW-SSIM values of all the testing slices. The FSIM and IW-SSIM values of the results after the processing of CNN are always greater than the ones before the processing. Table \ref{table_cr} presents the data compression ration in this experiment. It reaches 373.32 and indicates a tremendous data reduction.

The images in Figs.\ref{result3d_60} and \ref{result60} and the parameter values in Tables \ref{table_imagevalue}, \ref{table_evaluation} and \ref{table_cr}, demonstrate the validity of the sparse sampling, SVD compression, CNN restoration and the proposed lightweight solution of ICT for smart manufacturing. Within this solution, without significant image degradation, a drastic data reduction, a storage space save and an efficiency improvement are obtained.

\begin{table}[!t]
	% increase table row spacing, adjust to taste
	%\renewcommand{\arraystretch}{1.3}
	% if using array.sty, it might be a good idea to tweak the value of
	% \extrarowheight as needed to properly center the text within the cells
	\caption{The FSIM and IW-SSIM Values for image presented in Fig.\ref{result60}}
	\label{table_imagevalue}
	\centering
	% Some packages, such as MDW tools, offer better commands for making tables
	% than the plain LaTeX2e tabular which is used here.
	\begin{tabular}{c|c|c}
		\hline
		\multirow{2}{*}{FSIM} & \begin{tabular}[c]{@{}c@{}}Corrupted\\ CT Image\end{tabular}  &0.6575  \\ \cline{2-3}
		& \begin{tabular}[c]{@{}c@{}}Improved\\ CT Image\end{tabular}   & \textbf{0.9228}    \\ \hline
		\multirow{2}{*}{IW-SSIM} & \begin{tabular}[c]{@{}c@{}}Corrupted\\ CT Image\end{tabular}  &0.4019   \\ \cline{2-3}
		& \begin{tabular}[c]{@{}c@{}}Improved\\ CT Image\end{tabular}  &\textbf{ 0.9028}   \\ \hline
	\end{tabular}
\end{table}

\begin{table}[!t]
	% increase table row spacing, adjust to taste
	%\renewcommand{\arraystretch}{1.3}
	% if using array.sty, it might be a good idea to tweak the value of
	% \extrarowheight as needed to properly center the text within the cells
	\caption{The averaged FSIM and IW-SSIM values for all the 460 testing images}
	\label{table_evaluation}
	\centering
	% Some packages, such as MDW tools, offer better commands for making tables
	% than the plain LaTeX2e tabular which is used here.
	\begin{tabular}{c|c|c}
		\hline
		\multirow{2}{*}{FSIM} & \begin{tabular}[c]{@{}c@{}}Corrupted\\ CT Images\end{tabular}   & 0.7241   \\ \cline{2-3}
		& \begin{tabular}[c]{@{}c@{}}Improved\\ CT Images\end{tabular}  &\textbf{0.9258}  \\ \hline
		\multirow{2}{*}{IW-SSIM} & \begin{tabular}[c]{@{}c@{}}Corrupted\\ CT Images\end{tabular}  &0.4653   \\ \cline{2-3}
		& \begin{tabular}[c]{@{}c@{}}Improved\\ CT Images\end{tabular}   &\textbf{0.9006}    \\ \hline
	\end{tabular}
\end{table}

\begin{table}[!t]
	% increase table row spacing, adjust to taste
	%\renewcommand{\arraystretch}{1.3}
	% if using array.sty, it might be a good idea to tweak the value of
	% \extrarowheight as needed to properly center the text within the cells
	\caption{Compression Ratio with sparse sampling and SVD}
	\label{table_cr}
	\centering
	% Some packages, such as MDW tools, offer better commands for making tables
	% than the plain LaTeX2e tabular which is used here.
	\begin{tabular}{c|c|c|c}
		\hline
		& Sparse Sampling & SVD & Sparse Sampling and SVD \\ \hline\hline
	CR &\textbf{12}  &\textbf{31.11}  &\textbf{373.32}  \\ \hline
	\end{tabular}
\end{table}

\section{Conclusion}
In this paper, we proposed a lightweight solution of ICT devices for smart manufacturing with the support from IoT and CNN. Within this solution, ICT devices are separated into four function units. The data acquisition units remain in the production lines. Other three units are data storage, computing center and terminals and distributed by IoT. They are interconnecting and share instructions and data by IoT. It slims the ICT devices and enables the share of the same function units by the different production lines. This solution adopts sparse sampling to SVD compression technique to obtain a great data reduction. A compression ration about 300 can be achieved. It is helpful to improve the data transmission efficiency. The experimental results with auto components have demonstrate the validity of the proposed solution.

This solution depends on IoT and data compression techniques. Along with the develop of image processing and deep learning techniques, a greater data compression ratio can be obtained. Particularly, along with the application of the fifth generation of wireless communications technologies, IoT will have a revolutionary progress. They give a promising future to this solution.
% if have a single appendix:
%\appendix[Proof of the Zonklar Equations]
% or
%\appendix  % for no appendix heading
% do not use \section anymore after \appendix, only \section*
% is possibly needed

% use appendices with more than one appendix
% then use \section to start each appendix
% you must declare a \section before using any
% \subsection or using \label (\appendices by itself
% starts a section numbered zero.)
%

%\appendices
%\section{Proof of the First Zonklar Equation}
%Appendix one text goes here.
%
%% you can choose not to have a title for an appendix
%% if you want by leaving the argument blank
%\section{}
%Appendix two text goes here.

% use section* for acknowledgment
\section*{Acknowledgment}
We acknowledge support from the National Science and Technology Major Project of China (2018ZX04018001-006), National Natural Science Foundation of China (51975026), the Joint Fund of Research Utilizing Large-scale Scientific Facilities by National Natural Science Foundation of China and Chinese Academy of Science (U1932111).

% Can use something like this to put references on a page
% by themselves when using endfloat and the captionsoff option.
\ifCLASSOPTIONcaptionsoff
  \newpage
\fi

% trigger a \newpage just before the given reference
% number - used to balance the columns on the last page
% adjust value as needed - may need to be readjusted if
% the document is modified later
%\IEEEtriggeratref{8}
% The "triggered" command can be changed if desired:
%\IEEEtriggercmd{\enlargethispage{-5in}}

% references section

% can use a bibliography generated by BibTeX as a .bbl file
% BibTeX documentation can be easily obtained at:
% http://mirror.ctan.org/biblio/bibtex/contrib/doc/
% The IEEEtran BibTeX style support page is at:
% http://www.michaelshell.org/tex/ieeetran/bibtex/
\bibliographystyle{IEEEtran}
% argument is your BibTeX string definitions and bibliography database(s)
\bibliography{IEEEabrv,reference}
\end{document}